**Title:**

Deep-level transient spectroscopy of the charged defects in p-i-n perovskite solar cells induced by light-soaking

**Authors:**

Vasilev A.A.[1,2], Saranin D.S.[2*], Gostishchev P. A.[2], Tuhova M.P.[2] Didenko S.I[1,2], Polyakov A.Y.[1], and Di Carlo A.[3]

1. Department of Semiconductor Electronics and Semiconductor Physics, National university of Science & Technology MISIS, 4 Leninsky Ave., Moscow, 119049
2. Laboratory of Advanced Solar Energy (LASE), National university of Science & Technology MISIS, 6 Leninsky Ave., Moscow, 119049
3. Department of Electronic Engineering, University of Rome Tor Vergata, via del Politecnico 1, 100, 00133 Rome, Italy

\* Corresponding author Dr. Saranin.D.S.\* email: saranin.ds@misis.ru

**Highlights**

- The evaluation of the defect parameters of the p-i-n perovskite solar cells under light soaking stress.
- Analysis of impact of the charged defects on the performance and long-term stability of the $CsFAPbI_3$ based devices with Cl-doping.
- Cl-doping suppressed the formation of the antisite defect ($I_{Pb}$, $I_{FA}$) and iodine interstitials ($I_i$).
- Cl-doping increase efficiency and improve the light soaking stability.
- The presence of the defect vacancies ($V_I$, $V_{Cs}$) remains the critical factor for the long-term stable operation of the PSCs based on double cation compositions.

## Abstract

The long-term stability of halide perovskite solar cells (PSCs) remains the critical problem of this photovoltaic technology. Different structural defects formed in the thin-film perovskite films were considered as a main trigger for the decomposition of the absorber and corrosion of the interfaces in the device structure. The changes in the stability performance of the PSCs require a detailed analysis of the defects generated under external stress (light and heat). Using admittance, deep-level transient spectroscopy (DLTS) and reverse DLTS we determined the evolution of the defect energy levels in p-i-n PCS under continuous light soaking stress. We compared the impact of the charged defects on the performance and long-term stability of the $CsFAPbI_3$ based devices with and without Cl-doping. Despite the gain in the output performance of the PCSs, the devices with $CsFAPbI_{3-x}Cl_x$ showed improved light soaking stability. The $T_{80}$ (time required to reduce initial efficiency by 20%) for Cl-doped PSCs was 1280h, while for pure $CsFAPbI_3$ based devices only 650h. Three different defect energy levels were determined for different device configurations. We found that Cl-doping suppressed the formation of the antisite defects ($I_{Pb}$, $I_{FA}$) and iodine interstitials ($I_i$). The changes in the defect's energy levels after continuous light soaking stress were analyzed and discussed. The present work provides new insights for the defect behavior of PSCs under continuous external stress, revealing the physical-chemical impact of the Cl-additive strategy.

## Introduction

Over the past decade, the halide perovskite (HP) gained significant interest from the scientific community because of unprecedented progress in photovoltaics applications[1]. Perovskite solar cells (PSCs) showed rapid growth in the increment of power conversation efficiency(PCE), with a recent record at the level of 25.8%[2]. However, the degradation processes induced by compositional and structural defects in the bulk of thin-films of interfaces of the devices limit the long-term operation of PSCs with stable output performance[3]. Most promising chemical compositions of the HP-based light absorbers have a hybrid organic-inorganic nature. The general formula of the perovskite molecular structure is $ABX_3$, where A- cation is typically presented with methyl ammonium ($MA^+$), formamidine ($FA^+$) or cesium ($Cs^+$); B-cation is typically lead ($Pb^{2+}$), and X-anion is represented with halides - iodine ($I^-$), bromine ($Br^-$) or chlorine ($Cl^-$). PSCs could be fabricated with a low-cost solution processing methods[4]–inkjet printing[5], slot-die[6] or blade coating[7] which could significantly reduce capital expenditures of the industrial production[8]. On the other hand, using of solution-based crystallization methods at relatively low-temperatures induces the formation of different charged defects in microcrystalline HP-based absorbers. The defects in HP thin-films are associated to X-anion vacancies, A – and B- site cation interstitials, as well as with products of perovskite molecule decomposition – volatile $MA^+$ or $FA^+$; metallic lead $Pb^0$; Iodine ($I_2$); HI, etc [9]. Defects form "shallow" states (for example for iodine vacancies), or «deep» states (for examples for MA interstitials or free MA ions) which could form traps or recombination centers. The ionic defects in HP thin-films could trigger the electrochemical interaction at the interfaces with charge-transporting layers in PSCs, start the decomposition processes with presence of oxygen and moisture at the grain boundaries and contribute to the charge accumulation/depletion in the devices. The strategies for passivation and defect-healing for PSCs require specific methods for the rational identification of the defect-level parameters such as activation energy, cross-section, and concentration with quantitative assessment. Deep-level Transient Spectroscopy (DLTS) and Admittance Spectroscopy were widely regarded as standard techniques for the characterization of defects parameters in semiconductor-based device structures. The combination of both methods gives us an insight into the processes underlying the efficiency and stability of perovskite solar cells. Accessing deep-states through high-frequency capacitance transients or low-frequency conductivity is the shortest path to awareness of inner material properties such as charge carriers concentration, mobile ions distribution and built-in field screening.

The quantitative control of the defect parameters regarding the methods of the fabrication, doping levels and external stress factors is one of the key factors for the development of the stable performing solar cells. To date, the DLTS method is used more and more for the identification of defects in PSCs[10–13]. However, the evaluation of the defect characteristics in PSCs and its evolution with continuous influence of the stress factors (heat, high-intensity light) is still an open issue. Typically, the identification of the defect parameters performed only for "as fabricated" PSCs to distinguish the benefits of different crystallization methods[14], interface modification[15] etc.

In order to close this gap and shed light on the impact of accelerated stress tens on defect evolution, we made the attempt to analyze the changes of the defect parameters of PSCs under light soaking stress. We performed a DLTS investigation for p-i-n PSCs based on CsFAPbI$_3$ absorber with Cl-doping and analyzed the evolution of the defect parameters induced by external stress factor. We focus on the double cation HP with MA – and bromine free composition considering that multication HP is an effective strategy to improve photo and thermal stability of PSCs[16] and that CsFAPbI$_3$ engaged the interest of scientific community[17–20] thanks to the suppressed phase segregation[21] and thermal diffusion of the ionic defects[22].

We performed the stability tests following to the ISOS-L-2 protocol at open circuit conditions (V$_{oc}$) under continuous light soaking. The CsFAPbI$_{3-x}$Cl$_x$ based PSCs demonstrated higher stability with respect to CsFAPbI$_3$ with a T$_{80}$=1280h (T$_{80}$=time required to reduce initial efficiency by 20%) compared to T80=650h for CsFAPbI$_3$. The difference in the charged defect behavior before/after light soaking was evaluated by performing Admittance Spectroscopy and DLTS measurements with "classic" and reversed biasing conditions (RDLTS)[10,23]. We performed two DLTS sessions for the PCSs at the initial conditions and after 650h of stability tests. The results showed that fabrication of the pure triiodide CsFAPbI$_3$ tends to formation of the deep traps possibly related to the iodine ($V_I$) and cesium vacancies ($V_{Cs}$); antisites $I_{Pb}$ - $I_{FA}$ and iodine interstitials ($I_i$). Cl-doping suppressed the formation of the antistes and interstitials, even though $V_{FA}$ and $V_I$ were obtained. The presence of the $I_i$ after continuous light soaking in CsFAPbI$_3$ PCS was considered as one of the main instability factors.

## Results & Discussion

Perovskite solar cells were fabricated in the p-i-n planar configuration with the following architecture: Glass [1.1 mm]/ITO [330 nm]/NiO$_X$ [20 nm]/Perovskite/C$_{60}$ [40 nm]/Bathocuproine (BCP)[8 nm]/Cu [100 nm] encapsulated with a glass coverslip (see experimental for the details). The absorber layer was manufactured in two configurations – CsFAPbI$_3$ (Reference, here and after Reference sample) and CsFAPbI$_{3-x}$Cl$_x$ (Cl-doped, here and after Cl-sample). Briefly, the absorber films were fabricated with solution processing and using solvent engineering method for the crystallization. The molar ratio between Cs and FA cation was 1:4 and specific stoichiometry was Cs$_{0.2}$FA$_{0.8}$PbI$_3$. The Cl-doping was realized through the partial replacement of CsI used for the ink preparation to CsCl. The use of chlorine for anion substitution was considered as tool for the improvement of charge-carrier lifetimes [24,25] and diffusion lengths[26] in perovskite photoactive layers. The thickness of fabricated thin-films was 470 nm (measured with stylus profilometry) and didn't change after doping. We measured the performance of the fabricated PSCs under standard illumination conditions 1.5 AM G (Xe light simulator (AAA class)) (see fig.1a). Reference samples showed open circuit voltage (V$_{oc}$) values of 1.001 V, short-circuit current density (J$_{sc}$) of 22.5 mA/cm$^2$, filling factor (FF) equal to 0.755 and PCE=17.06%. The Cl-doping boosts the performances of PSC with an increase of V$_{oc}$ up to 1.012 V(+ 1% with respect to undoped device), J$_{sc}$ up to 25.2 mA/cm$^2$(+12%) and PCE up to 19.39%(+13.6%). Following to the initial characterization, we set up the stability tests according to the ISOS-L-2[27] with a continuous light soaking at V$_{oc}$ conditions (see experimental for the details). The temperature of the samples was 63.5±1.5 °C under light exposure. The accelerated stress induced a degradation of the PSC characteristics as shown in Fig.**1b-d**. However, the T$_{80}$ period for the reference sample was equal to 650h, while with the Cl-doping T$_{80}$ increased up to 1280h (Fig. 1d).

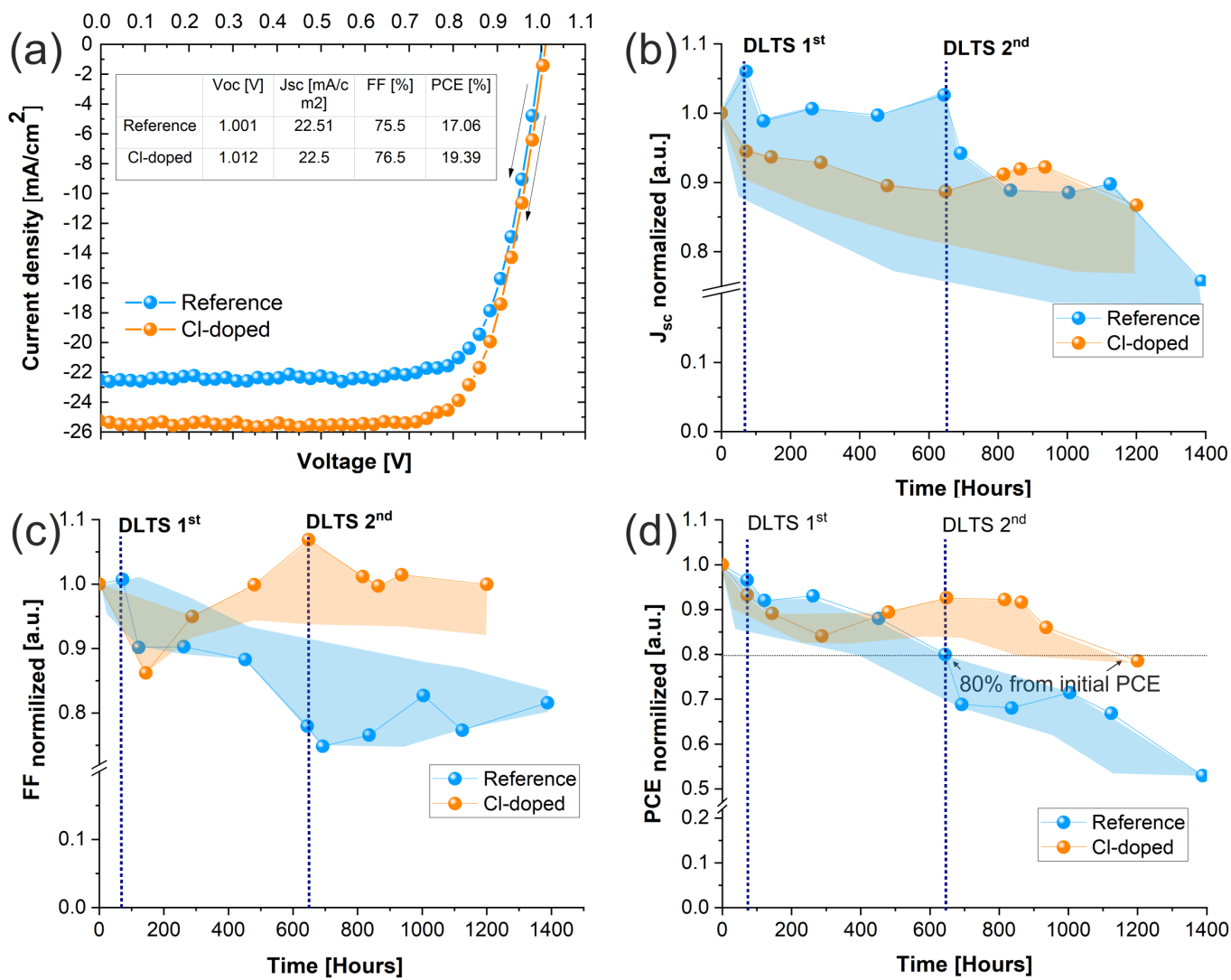

Figure 1 – Output JV performance of fabricated PSCs based on reference and Cl-doped absorbers measured under standard conditions 1.5 AM G (a) Normalized output parameters of the fabricated PSCs under continuous light soaking stress – $J_{sc}$ (b), FF(c), PCE (d). The symbols represented the results for the most stable devices, the shaded area she range of the experimental data measured for all investigated PSCs.

For both configurations, the major impact on the reduction of the PCE was given by the decrease of $J_{sc}$ and FF while $V_{oc}$ was more stable (see Fig. **S1**). In parallel to the stability tests, we performed two sessions of the DLTS measurements to determine the evolution of the defects induced by the external stress which was the main aim of our investigation. The first set of DLTS was performed after 72h of stability tests and the second one was done after 650h (i.e. at the T80 of reference device.

## Capacitance-Voltage profiling

Capacitance-frequency measurements on reference and Cl- samples were carried out to to measure the density of uncompensated shallow donor and define measurement frequencies for further DLTS routines. All C-F profiles at room temperature have droop for frequencies lower than 200 Hz because of mobile ions screening and droop higher than 200 kHz because of the series resistance of samples. The plateau region between those droops is determined by the thickness of the perovskite layer and stays in good agreement with profilometer data (Fig. **S2**).

In the C-V characteristics (Fig. **S3**), measured on the plateau frequency, the capacitance has been constant from reverse to a small forward bias, which is associated with the complete depletion of the perovskite active layer. Donor concentration was determined from data at sufficiently large forward biases (~0.5 V), such that the forward current and diffusion capacitance did not yet prevent correct measurements of the p-i-n diode capacitance associated with a change in the thickness of the volume charge layer. In such conditions, $1/C^2$ vs V data was linear, and it was possible to determine the concentration of uncompensated shallow donors (~$10^{16}$ cm$^{-3}$) and the cut-off voltage of 0.9–1.2 V, which is close to the $V_{oc}$ voltage for all the samples.

## Admittance Spectroscopy

Characteristics of deep-levels and mobile ions were determined using Admittance Spectroscopy and DLTS techniques. Measurements of admittance were carried out using liquid nitrogen cryostat at 200–350 K temperature range within 20 Hz–2 MHz frequency band using LCR-meter Keysight E4980A. Admittance data were recorded at a ~3 K/min cooling rate with temperature precision better than 0.1 K and plotted in $G/\omega$ vs $T$ coordinates (being $G/\omega$ the AC conductance $G$, normalized to the angular frequency ω). From low to high temperatures, a low-frequency droop appeared in the capacitance and a peak appeared in the frequency dependence of the AC conductivity, because of the relaxation time for the process causing a large capacitance became equal to the period of the test voltage oscillation ($\tau \cdot \omega = 1$). Measurements of admittance spectra for these frequencies made it possible to calculate the temperature dependence of the relaxation time and find the characteristic activation energy of the process, which turned out to be 0.41–0.45 eV for all samples and is associated with a change in the charge of mobile ions, as illustrated in Figure **2**.

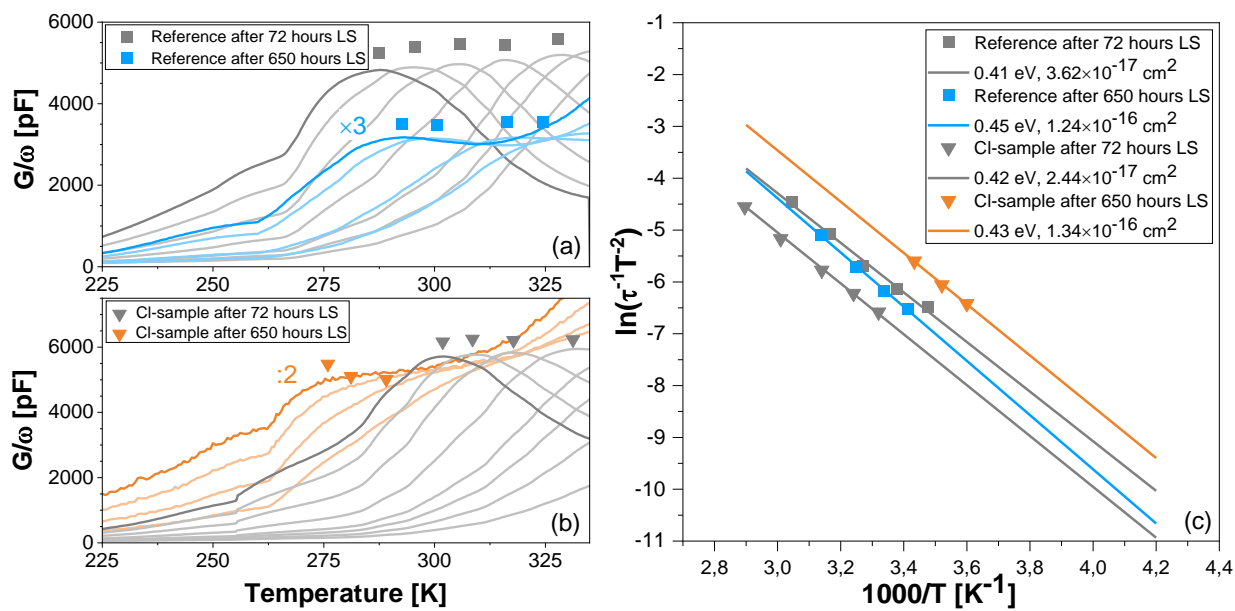

Figure 2 – Admittance spectra recorded at zero bias for (a) Reference and (b) Cl- sample. (c) Activation energy evaluation from Arrhenius plot

Density determination of those 0.41–0.45 eV ions is limited due to the screening of the probing signal deeper than $\lambda_D = 0.25$ μm (from 20 nF at 350 K at 20 Hz) by near-surface mobile ion charge [3].

### Deep-Level Transient Spectroscopy

The same measurement setup was used to perform DLTS. In this measurement, device under test is constantly biased at -0.1 V and capacitance transient, *C(t)* is recorded after fast 50 ms forward +0.5 V pulse. This transient contains information about prevalent processes at certain temperatures with deep-levels and/or mobile ions involved. After 200-350 K temperature sweep data is plotted in $(C(t_2) - C(t_1))/C(t_\infty)$ vs $T$ coordinates, where peak position corresponds to relaxation time $\tau = (t_2 - t_1)/\ln(t_2/t_1)$ (Fig. 3c,d). Peak temperatures and corresponding relaxation times are then plotted in $\ln(\tau^{-1}T^{-2})$ vs $1/T$ Arrhenius plot with activation energy extraction from data fitting (see supplementary materials, Fig. S4).

Recording transients with reversed biasing conditions (+0.5 V bias and -0.1 V pulse) we would perform the so-called Reverse-DLTS (RDLTS)[28,29]. It should be noted, that in conventional semiconductors, peaks in the reverse DLTS is an exception and are always associated with some secondary phenomena, such as the presence of a barrier for capture by deep-levels or a change in the threshold voltage of Field Emitting Transistors (FETs)[30] as a result of charge capture by deep-levels under the gate (Fig. **3a,b**).

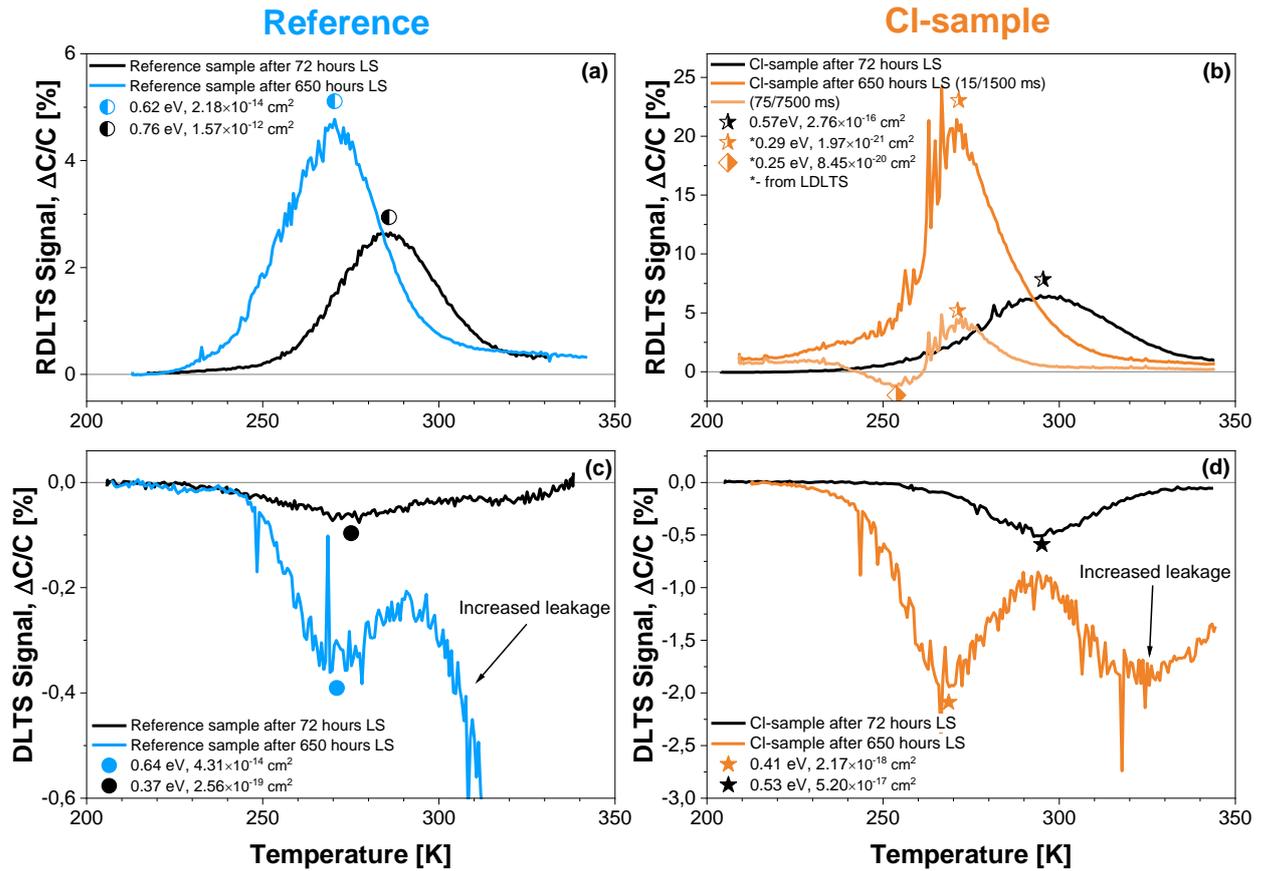

Figure 3 – (a), (b) RDLTS and (c), (d) DLTS spectra for R- and Cl-samples respectively before (black lines) and after (colored lines) Light-Soaking.

The DLTS spectra exhibited a peak with a low amplitude and characteristic activation energy of 0.37 eV for the reference sample after 72 hours of light soaking, which is close to the activation energy calculated from the admittance spectra.

A peak with a large amplitude and activation energy of 0.76 eV was found in the RDLTS (Fig 3a,c, black lines). In the perovskite p-i-n diodes studied in our previous works [10,15,28], we associated the peaks in RDLTS to the rearrangement of the space charge caused by the movement of mobile ions. Usually, in this case, a peak in RDLTS corresponds to a "mirror" peak with similar activation energy, but with opposite sign in DLTS. In the samples studied in this work, the peak amplitude in DLTS is much smaller than in RDLTS. Apparently, by varying the voltage polarities applied to the sample, the efficiency of moving ions is different, which may be due to the different distance they must go to the corresponding contact.

In the DLTS signal of the Cl-sample, the amplitude of the main peak noticeably increased after long exposure, and the corresponding activation energy decreased (Fig 3d). In the RDLTS spectra, the amplitude of the peak mirroring the DLTS peak increased significantly with a shift to lower temperatures, like the DLTS peak after light soaking long exposure. Moreover, an additional peak of opposite polarity appeared (Fig. 3b).

To separate these two different sign peaks appeared in Cl-sample after 650 hours of light-soaking we used numerical Laplace transform routines. Our approach based on L1 regularization method using FISTA algorithm [31] for finding sparse regularized solution $f(s)$ of inverse Laplace transform:

$$F(t) = \int_0^\infty f(s)e^{-st}\,ds \quad (1)$$

$$f(s): \min_f \{\|Lf - F\|_2^2 - \alpha\|f\|_1\} \quad (2)$$

Here $\|Lf - F\|_2^2$ – denotes mean-square error norm, and $\alpha\|f\|_1$ – L1 penalty. By adding the last term, we replace the initial ill-conditioned problem with a "nearby" well-conditioned whose solution approaches the desired solution. The optimal regularization parameter for every processed transient is determined using L-curve criteria [32]. This allows us to deconvolute two peaks, appeared in Cl-sample RDLTS after 650 hours exposure and fit the data. Results are presented in Figure 4.

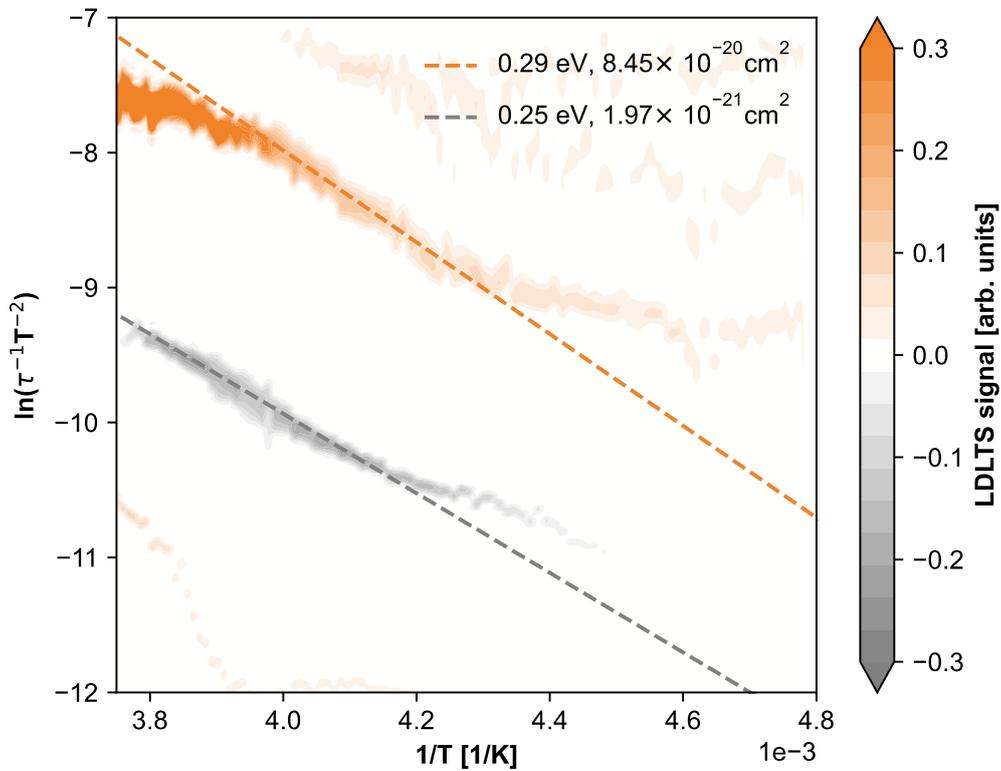

Figure 4– Arrhenius plot for Cl-sample in RDLTS regime with two centers fitted with 0.25 eV (grey line) and 0.29 eV (orange line)

Resulted activation energies for negative and positive amplitude peaks in Cl-sample RDLTS (Fig. 3b, orange lines) are 0.25 eV and 0.29 eV, respectively.

Data obtained with Admittance, DLTS, RDLTS measurements are summarized in Table 1 and Figure 5. Defect inducing rates visualized in **Fig. S4** in supplementary materials and represents the correlation between concentration of the traps with LS time.

Table 1 – Traps concentrations and activation energies for R- and Cl-samples before and after Light-Soaking (* – positive amplitude center from Fig. 3b, half-filled orange star)

| Sample | Method | 72 hours Light-Soaking | | | 650 hours Light-Soaking | | |
|---|---|---|---|---|---|---|---|
| | | Concentration ($cm^{-3}$) | Activation energy (eV) | Possible origin [Link] | Concentration ($cm^{-3}$) | Activation energy (eV) | Possible origin [Link] |
| Reference sample | Admittance | – | 0.41 | $V_I$ [33] | – | 0.45 | $V_I$ [33] |
| | DLTS | $4.33 \cdot 10^{12}$ | 0.37 | $I_{FA}$[34] or $V_{Cs}$[35] | $1.39 \cdot 10^{13}$ | 0.64 | $V_I$ or $I_i$ |
| | RDLTS | $1.25 \cdot 10^{14}$ | 0.76 | $I_{Pb}$32] or $I_i$[35] | $1.93 \cdot 10^{14}$ | 0.62 | $V_I$ or $I_i$ |
| Cl-sample | Admittance | – | 0.42 | $V_I$ [33] | – | 0.43 | $V_I$ [33] |
| | DLTS | $5.63 \cdot 10^{13}$ | 0.53 | $V_{FA}$ [34,36] | $2.93 \cdot 10^{14}$ | 0.41 | $V_I$ [33] |
| | RDLTS | $2.13 \cdot 10^{14}$ | 0.57 | $V_{FA}$ [34,36] | *$9.44 \cdot 10^{14}$ | *0.29 | $V_{Cs}$[35] |

The calculated defect parameters presented in the Tab.1 requires a thorough discussion. In fact, our multi-cation compositions require to consider the specifics behavior of both $FAPbI_3$ and $CsPbI_3$ perovskites.

Despite the improvement in PCE for Cl-samples compared to the references, the traps concentration ($N_t$) extracted from DTLS is $5.63 \cdot 10^{13}$ $cm^{-3}$. For not doped PSCs, the $N_t$ value was approximately ten times smaller - $4.33 \cdot 10^{12}$ $cm^{-3}$. At the same time, the RDLTS data show very close values of $N_t$ in the order of $10^{14}$ $cm^{-3}$ for reference and Cl- samples. The activation energy ($E_A$) of the defects extracted from the Admittance spectroscopy showed almost equal results for reference and Cl-doped perovskites with $E_A$ value of 0.41 and 0.45 eV, respectively. According to the calculated energy levels, the ~0.4 eV trap can be associated to iodine vacancy ($V_I$) related to the migration of the defect in axial-to-equatorial direction of $CsPbI_3$ molecular structure[33]. The $E_A$ values for the reference sample extracted from DLTS and RDLTS displayed acceptor defects at 0.37 eV and 0.76 eV. The 0.37 eV trap can be indicated as iodine - formamidinium antisites ($I_{FA}$)[34] or cesium vacancy($V_{Cs}$)[35]. The 0.76 trap can be related to the iodine - lead antisites ($I_{Pb}$) [34] or iodine interstitials ($I_i$)[35]. The deep traps of 0.53 and 0.57 eV measured for the Cl-samples were determined as formamidinium vacancies ($V_{FA}$)[34,36]. Clearly, incorporating the Cl-anion significantly changed the defect behavior. According to the observed data, anion engineering suppressed the formation of the antisite defect energy levels. The impact of the Cl on defect energy of halide perovskite absorbers fabricated with different crystallization techniques were already reported [37]. *Tan et al.* demonstrated[38] that the formation energy of Pb–Cl antisite defects is much greater than that of a Pb–I antisite, showing that antisite defects are suppressed by the presence of interfacial Cl atoms.

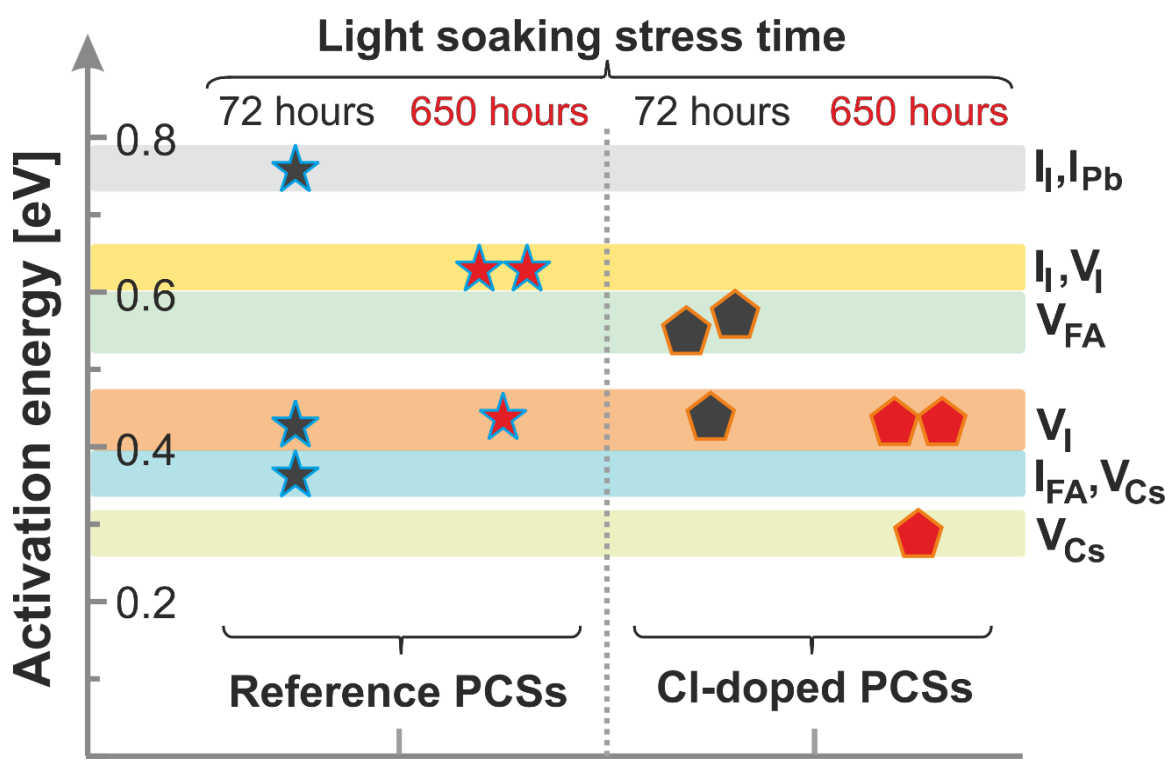

Figure 5 – The evolution of the activation energy for defects measured by admittance, DLTS, RDLTS for the reference (stars) and Cl-doped (hexagons) PSCs at the beginning (72h, black symbols) and at the end (650 h, red symbols) of light soaking stress. Labelling of the levels is provided on the right and refers to the data and references reported in Table 1

After 650h of light soaking, the defects parameters in PSCs showed an interesting evolution. The traps concentration extracted by DLTS showed approximately the same magnitude of increase for both studied device configurations. The $E_A$ values for the defect energy levels calculated from the admittance spectra showed a shift of +0.04 eV for reference samples and +0.01 eV for Cl-samples. Despite this, the $E_A$ extracted from DLTS after LS was 0.64 eV very different from the 0.37 eV obtained at 72 hours of LS. The 0.64 eV trap can be associated to the migration energy for the iodine vacancy ($V_I$) in equatorial axis of $CsPbI_3$ [33] or iodine interstitial($I_i$)[34,35], which are both related to the X-site atom position. Approximately the same $E_A$ of 0.62 eV was obtained after 650h of LS in reference samples with the RDLTS. In contrast, the $E_A$ of the defect for Cl-samples was reducing after prolonged LS. We obtained a $E_A$= 0.41 eV, which could be indicated as iodine vacancy ($V_I$) in agreement to the admittance data. The results from RDLTS measurements showed even a smaller $E_A$=0.29 eV. This shallow energy level could be related to the migration of the $V_{Cs}$[35].

The losses in the performance of the PSCs is mostly related to the decomposition processes of the perovskite absorber and migration of the point defects [34],[31]. The imperfections and point defects act as a triggers for the decomposition of the halide perovskite absorbers and degradation of the interfaces in the device[40]. The presence of the negatively charged antisites - $I_{Pb}$ and $I_{FA}$, observed only for the reference samples after 72h of light soaking, can been related to deep non-radiative recombination centers[41], that migrate towards the metal electrode and initiate the corrosion processes[42]. This explains the lower PCE for $CsFAPbI_3$ PSCs compared to the $CsFAPbI_{3-x}Cl_x$ due to the trapping of the charge carriers. Also, It was found that formation of the iodine interstitial in the formamidinium based perovskite absorbers accelerates the transition from the cubic photoactive phase to the tetragonal inactive phase[43] and induces the formation of the neutral $I_2$ from two filled traps[44]. The presence of $I_i$ energy levels in the reference $CsFAPbI_3$ PCSs was always observed during the measurements and could be considered as one of the major instability factors. According to the observed results, the Cl-anion substitution in the chemical composition of the absorber in the Cl-doped $CsFAPbI_3$ perovskite suppress the formation of the deep recombination centers represented by antisites and enhanced the phase stability due to the reduced impact of the interstitials defects. Thisin turn, permit a structural stabilization of the $CsFAPbI_3$ and improves charge carrier transport in the devices. On the other hand, for both device configurations we found defect levels related to $V_I$ or $V_{Cs}$, which could trigger the reaction of the photogenerated carriers with iodine sub-lattice and formation of $I_2$[45] or CsI. Therefore, the Cl-additive strategy for the crystallization of the multi-cation perovskite absorbers requires further improvement to overcome the problems with the formation of the structural defects which lead to the structural instability under external stress. The new attempts should include precise chemical composition to exclude the impact of the side impurities, more advanced crystallization method for uniform nucleation and full compliance with the stoichiometry of the chemical composition.

## Conclusions

Using Admittance Spectroscopy, DLTS and RDLTS measurements, we evaluated the changes in the defect parameters for the $CsFAPbI_3$ PSCs with and without Cl-doping under continuous light soaking stress. Partial Cl-anion substitution in the chemical composition of the perovskite absorber improved the (maximum) PCE value from 17.06% to 19.39% and the light soaking stability at $V_{oc}$ conditions from $T_{80}$=650h for reference sample ($CsFAPbI_3$) to 1280h for Cl-doped devices. We found three kinds of the charged defects in PSCs based on stoichiometric triiodide perovskite - $CsFAPbI_3$: 0.41 eV extracted from Admittance (possibly $V_I$,); 0.37 eV extracted from DLTS (possibly $I_{FA}$ or $V_{Cs}$,) and 0.76 eV extracted from RDLTS (possibly $I_{Pb}$ or $I_i$,). The defects for Cl-doped PSCs showed different energy levels: 0.42 eV extracted from Admittance (possibly $V_I$); 0.53 and 0.57 eV (possibly $V_{FA}$ for both), extracted from DLTS and RDLTS, respectively. We conclude that elimination of the $I_{Pb}$ antistites and iodine interstitials ($I_i$) in the Cl-doped PSCs significantly reduced the impact of the nonradiative recombination to the device operation and improved the efficiency of the charge transport.

Interestingly, that after 650h of the continuous light soaking, the defect energy levels calculated from the DLTS and RDLTS were significantly shifted in both device configurations. After external stress, the $CsFAPbI_3$ based PSCs showed traps at 0.62 and 0.64 eV (possibly $V_I$ or $I_i$). For $CsFAPbI_{3-x}Cl_x$ based devices, the activation energy values became shallower – 0.41 eV (possibly $V_I$) and 0.29 eV (possibly $V_{Cs}$). We found that Cl-doping suppressed the formation of the antisite defect ($I_{Pb}$, $I_{FA}$) and iodine interstitials ($I_i$), that could trigger unfavorable phase transitions and degradation of the interfaces. Our investigation clearly showed that Cl-anion substitution changes the mechanisms of the defect formation and delays the dynamics of structural degradation in PCSs. The presence of the defect vacancies ($V_I$, $V_{Cs}$) remains the critical factor for the long-term stable operation of the PSCs based on double cation compositions. The present work provides new insights for the defect behavior of PSCs under continuous external stress, reveals the critical points for the Cl-additive strategy and stabilization of the device performance.

# Appendix

*Experimental section*

*Materials*

Herein, the organic solvents dimethylformamide (DMF), N-Methyl-2-pyrrolidone (NMP), chlorobenzene (CB), 2-propanol (IPA), and methoxyethanol (MOE) were used in anhydrous, ultra-pure grade from Sigma Aldrich (Germany). The device was fabricated on $In_2O_3:SnO_2$ (ITO, 10 ohm/sq) coated glass from Kaivo. We used acidified nickel chloride ($NiCl_2$) as a precursor for the NiO hole-transporting layer (HTL). Lead Iodide (PbI2 99.9 % purity LLC Chemosynthesis (Russia)) and cesium iodide (CsI 99,9 %, Lanhit (Russia)) formamidinium iodide (FAI, 99.99 % purity from GreatcellSolar) were used to prepare the perovskite ink. Fullerene (C60 99.9 % MST, Latvia) and bathocuproine (BCP, >99.5 % sublimed grade, Osilla Inc., UK) were used as an electron-transporting layer (ETL).

*PSC fabrication*

The p-i-n-structured solar cell with the following stack ITO/$NiO_x$/$CsFAPbI_3$/C60/BCP/Cu was fabricated via an optimized route. Firstly, the ITO substrates were cleaned with detergent, de-ionized water, acetone, and IPA in an ultrasonic bath. Then the substrates were activated under UV-ozone irradiation for 30 min. The solution of nickel nitrates 0.15 M in MOE for NiO HTL was spin-coated at 4000 RPMs (30 s), dried at 120 °C (10 min), and annealed at 300 °C (1 h) in ambient atmosphere. The $CsFAPbI_3$ film was crystallized on top of HTL with a solvent-engineering method. The perovskite precursor was spin-coated with the following ramp: (1s – 1000 rpm, 4 sec – 3000 rpm / 30 sec – 5500 rpm). In details, 420 µL of CB were dropped on the substrate on the 10$^{th}$ second after starting the first rotation step. Then the substrates were annealed at 70 °C (1 min) and 105°C (30 min) to form the appropriate perovskite phase. C60 was deposited with the thermal evaporation method at $10^{-6}$ Torr vacuum level. The free BCP interlayer was spin-coated at 4000 RPMs (30 s) and annealed at 50 °C (1 min) for the reference devices. The copper cathode was also deposited with thermal evaporation through a shadow mask to form a 0.14 cm$^2$ active area for each pixel. All fabricated devices were encapsulated with a UV curable epoxy from Osilla Inc. and a glass coverslip

*The Device characterization methods*

The photovoltaic (PV) parameters for the fabricated PSCs (open-circuit voltage - Voc, short circuit current density - Jsc, fill factor – FF, and power conversation efficiency – PCE) were calculated and statistically analyzed from the IV characteristics measured under standard conditions of 1.5 AM G light spectrum. We used Xenon arc-lamp based solar simulator (ABET 3000) with AAA grade of conformity to the reference Air Mass 1.5 Global spectrum of the terrestrial solar light (1.5 AM G standard - (ASTM) G-173). The standard power density of the incoming light (100 mW/cm2) was calibrated with certified Si- reference solar cell. The scan sweeps and the MPPT tuning measurements were performed with Keithley 2400 SMU.

The stability tests were performed with an LED light source (4000 K) equilibrated to the 1 sun conditions with alignment to $J_{sc}$ values. The device were placed under light-soaking condition with open-circuit voltage regime. The stability was estimated through the periodical JV measurements every 100 – 200 hours.

*CV profiling, Admittance and Deep-level transient spectroscopy*

C-F and C-V profiling were performed using Keysight Precision LCR Meter E4980A for accessing material thickness and carrier density profiles. Admittance Spectroscopy and DLTS were carried out within the 200÷350K temperature range in liquid nitrogen cryostat LN-120 from CryoTrade Engineering. Transient analysis for Laplace DLTS routines with the following peak deconvolution for this work was executed in Python 3 using FISTA [31] algorithm. This code is hosted on GitHub (https://github.com/nocliper/ilt)